\documentclass[12pt]{article}
\usepackage{graphicx,epsfig}
\usepackage{palatino}
\def\beq{\begin{equation}}
\def\eeq{\end{equation}}
\def\bea{\begin{eqnarray}}
\def\eea{\end{eqnarray}}
\def\bq{\begin{quote}}
\def\eq{\end{quote}}

\def\bq{\begin{quote}}
\def\eq{\end{quote}}

\begin{document} 
\baselineskip 18pt 
\vspace*{-1in} 
\renewcommand{\thefootnote}{\fnsymbol{footnote}} 
\begin{flushright} 
hep-ph/yymmnnn\\
TIFR/TH/07-24\\
\end{flushright} 
\vskip 65pt 
\begin{center} 
{\Large \bf \boldmath Associated production of a Kaluza-Klein excitation of a 
gluon with a $t \bar t$ pair at the LHC}\\
\vspace{8mm} 
{\large\bf M. Guchait$^{(1)}$\footnote{guchait@tifr.res.in}, 
F. Mahmoudi$^{(2)}$\footnote{nazila.mahmoudi@tsl.uu.se} and
         K.~Sridhar$^{(3)}$\footnote{sridhar@theory.tifr.res.in}
}\\ 
\vspace{10pt} 
\end{center}
{\it 1. Department of High Energy Physics, 
                     Tata Institute of Fundamental Research,  
                     Homi Bhabha Road, 
                     Bombay 400 005, India.\\ } 
{\it 2. High Energy Physics, Uppsala University,\\ 
Box 535, 75121 Uppsala, Sweden.\\ } 
{\it 3. Department of Theoretical Physics, 
                     Tata Institute of Fundamental Research,  
                     Homi Bhabha Road, 
                     Bombay 400 005, India.\\ } 

\normalsize
 
\vspace{20pt} 
\begin{center} 
{\bf ABSTRACT} 
\end{center} 

\noindent In Randall-Sundrum models, the Kaluza-Klein (KK) excitations
of the gluon, $g_{KK}$ have enhanced couplings to the right-handed quarks. In
the absence of a $gg g_{KK}$ coupling in these models, the single production
of a $g_{KK}$ from an initial $gg$ state is not possible. The search for
other production mechanisms at the LHC, therefore, becomes important. We
suggest that the associated production of a $g_{KK}$ with a $t \bar t$
pair is such a mechanism. Our study shows that through this process the
LHC can probe KK gluon masses in the range of 2.8 -- 2.9 TeV. 

\vskip12pt 
\noindent 
\setcounter{footnote}{0} 
\renewcommand{\thefootnote}{\arabic{footnote}} 
 
\vfill 
\clearpage 

\setcounter{page}{1} 
\pagestyle{plain}
\baselineskip 19pt 


\noindent The Randall-Sundrum (RS) model \cite{rs} is a five-dimensional model
with the fifth dimension $\phi$ compactified on a ${\bf
S}^1/{\bf Z^2}$ orbifold. The compactification radius $R_c$ 
is somewhat larger than $M_P^{-1}$, the Planck length. 
The fifth dimension is a slice of anti-de Sitter spacetime
and is strongly curved. At the fixed points
$\phi=0,\ \pi$ of the orbifold, two D-3 branes are located and
are known as the Planck brane and the TeV brane, respectively.
The Standard Model fields are localised on the TeV brane 
while gravitons exist in the full five-dimensional spacetime.
The five-dimensional spacetime metric is of the form
\begin{equation}
ds^2 = e^{-{\cal K}R_c\phi}\eta_{\mu\nu}dx^{\mu}dx^{\nu}~+~R_c^2d\phi^2 .
\end{equation} 
Here ${\cal K}$ is a mass scale related to the curvature. 
The factor ${\rm exp}(-{\cal K} R_c\phi)$ is called the
warp factor and serves to rescale
masses of fields localised on the TeV-brane. For example,
$M_P=10^{19}$ GeV for the Planck brane at $\phi=0$ gets rescaled to 
$M_P {\rm exp}(-{\cal K} R_c\pi)$ for the TeV brane at $\phi=\pi$. 
The warp factor generates $\frac{M_P}{M_{EW}} \sim 10^{15}$ 
by an exponent of order 30 and solves the hierarchy problem. 
For this mechanism to work, one will have to ensure that the
radius $R_c$ is stabilised against 
quantum fluctuations  and this can be done by introducing a bulk 
scalar field which generates a potential that allows for
the stabilisation \cite{gold}. The model predicts a discrete
spectrum of Kaluza-Klein (KK) excitations of the graviton and
these couple to the Standard Model fields with a coupling
that is enhanced by the warp factor to be of the order of electroweak
strength. Several collider implications of these gravitons
resonances have been studied in the literature \cite{pheno}. 

The AdS/CFT correspondence \cite{maldacena} allows us to get an
understanding of the RS model in terms of a dual theory -- a strongly
coupled gauge theory in four dimensions \cite{holography}. This
four-dimensional theory is conformal all the way from the
Planck scale down to the TeV scale and it is only the presence
of the TeV brane that breaks the conformal symmetry.
The KK excitations as well as the fields localised on the TeV brane are
TeV-scale composites of the strongly interacting theory. Since in the
RS model, all SM fields are localised on the TeV brane, the AdS/CFT
correspondence tells us that the RS model
is dual to a theory of TeV-scale compositeness of the entire SM. 
Such a composite theory is clearly unviable: but is there a way
out? There seems to be -- and the 
simplest possibility is to modify the
model so that only the Higgs field is localised on the TeV brane
while the rest of the SM fields are in the bulk \cite{pomarol}.

Flavour hierarchy, consistency with electroweak precision
tests and avoidance of flavour-changing neutral currents can be used 
as guiding principles in constructing such models \cite{models}. 
In particular, in order to avoid an unacceptably large contribution
to the electroweak $T$ parameter an enhanced symmetry in the bulk
like $SU(2)_L \times SU(2)_R \times U(1)_{(B-L)}$ may be used. The
heavier fermions need to be closer to the TeV brane so as
to get a large Yukawa coupling i.e. overlap with the Higgs.
In other words, the profiles of the heavier fermions need
to be peaked closer to the TeV-brane. Conversely, 
the fermions close to the Planck brane will have small Yukawa couplings.
However, while the large Yukawa of
the top demands proximity to the TeV brane, the left-handed
electroweak doublet, $(t, b)_L$, cannot be close to the TeV brane
because that induces non-universal couplings of the $b_L$ to the
Z constrained by $Z \rightarrow b \bar b$. So the doublet needs to be  
as far away from the TeV brane as allowed by $R_b$ whereas the $t_R$
needs to be localised close to the TeV brane to account for the
large Yukawa of the top. We stress that this is one model realisation;
a different profile results, for example, in models that invoke 
other symmetry groups in the bulk \cite{alternates}. 
It has been found that in order to avoid huge effects of flavour-changing
neutral currents (FCNCs) and to be consistent with precision tests of 
the electroweak sector, the masses of the KK modes of the gauge
bosons have to be strongly constrained. The resulting 
bounds on the masses of the
KK gauge bosons are found to be in the region of 2-3 TeV \cite{models} 
though this bound
can be relaxed by enforcing additional symmetries. A review of the
literature on this subject can be found in Ref.~\cite{review}.

The collider implications of this scenario has been studied
recently \cite{studies}.
While some of these studies have focussed on graviton
production \cite{graviton},
the interesting signals for this scenario is the production
of KK gauge bosons and, for the LHC in particular, the production
of KK gluons. The KK gluon couples strongly to the $t_R$, with a
strength which is enhanced by a factor $\xi$ compared to the QCD coupling
where $\xi\equiv \sqrt{{\rm log} (M_{pl}/{\rm TeV})} \sim 5$. 
Consequently, it decays predominantly to tops if produced. To the left-handed
third-generation quarks, the KK gluon couples with the same
strength as the QCD coupling whereas to the light quarks its couplings 
are suppressed by a factor $1/\xi$. The problem in producing the KK
gluon at a collider, however, is that its coupling to two gluons 
vanishes because of the orthogonality of the profiles of these particles
and, therefore, the gluon production mechanism at a hadron collider
cannot produce the KK gluon at leading order.
The KK gluon can, therefore, be produced by annihilation of light
quarks and this production mechanism has been studied in the
context of the LHC \cite{kkgluon}. The same mechanism has also been
studied in the context of Tevatron to derive a model-independent
bound of 770 GeV from the Tevatron top cross-section \cite{us}.

In this paper, we study the production of KK gluons in association
with a $t \bar t$ pair. A similar process has been recently discussed
in Ref.~\cite{abdel}. In this process the $t \bar t$
pair can be produced from both $gg$ and $q \bar q$ initial states through
the usual QCD processes and the KK gluon, $g_{KK}$ can then be radiated from 
one of the heavy-quark legs. The fact that the $gg$ initial state contributes
to the associated production process makes it appealing. Also 
the process directly probes the coupling of the $g_{KK}$ to the tops
which is an important feature of the new dynamics.

The Feynman diagrams for the $q \bar q \rightarrow g_{KK} t \bar t$ and the 
$gg \rightarrow g_{KK} t \bar t$ subprocess are shown
in Fig. 1. We have computed the matrix elements for 
these subprocesses using FORM. The $g_{KK}$ is produced
on-shell and we ignore virtual effects. The produced $g_{KK}$ decays
into a $t \bar t$ pair yielding two pairs of $t \bar t$ in the final
state. The background to this signal of two non-resonant $t \bar t$ pairs
coming from QCD processes has been computed using ALPGEN \cite{alpgen}.
The squared-matrix elements for the signal are available in the form of a
Fortran code but the expressions are too lengthy to reproduce here. 
A KK gluon with a mass just a little above
the $t \bar t$ threshold has a very large branching into top pairs: the
branching ratio is about 92.5\% \cite{kkgluon}. Since we are interested
in KK gluon masses well above the $t \bar t$ threshold we will assume that
the produced $g_{KK}$ decays with this branching ratio into a $t \bar t$
pair.



\begin{figure}[!h]
\hbox{(a)} \hbox{\includegraphics[scale=0.3]{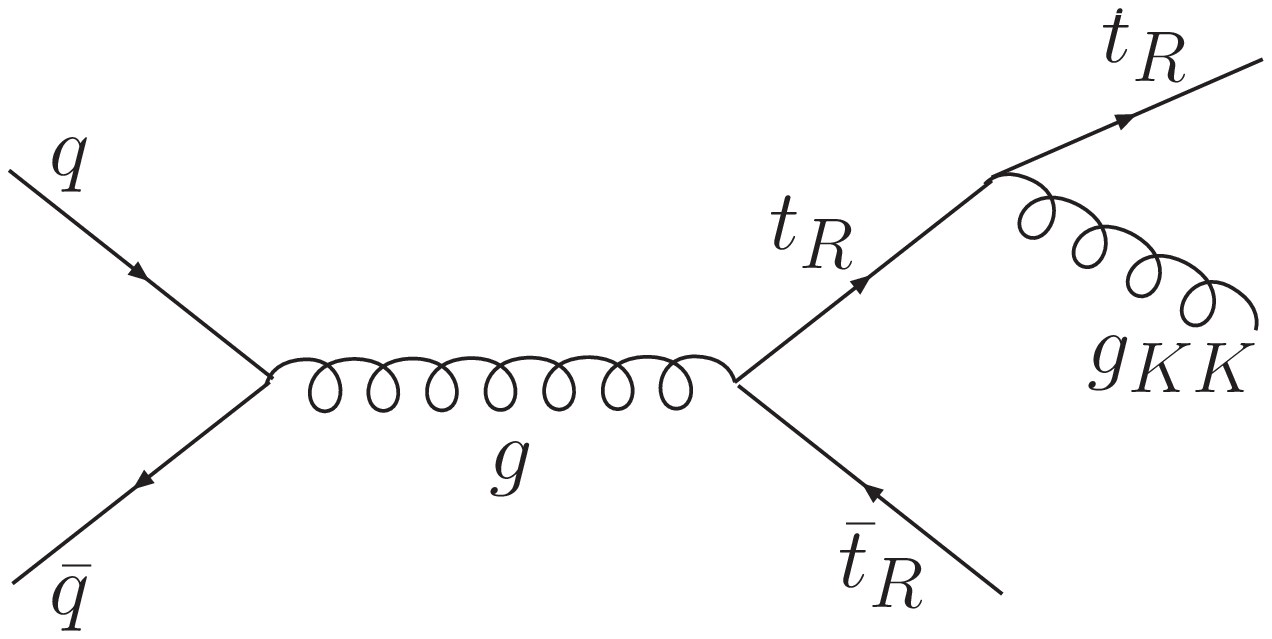}}\vspace{-1.6cm}\hspace{5cm}\hbox{\includegraphics[scale=0.3]{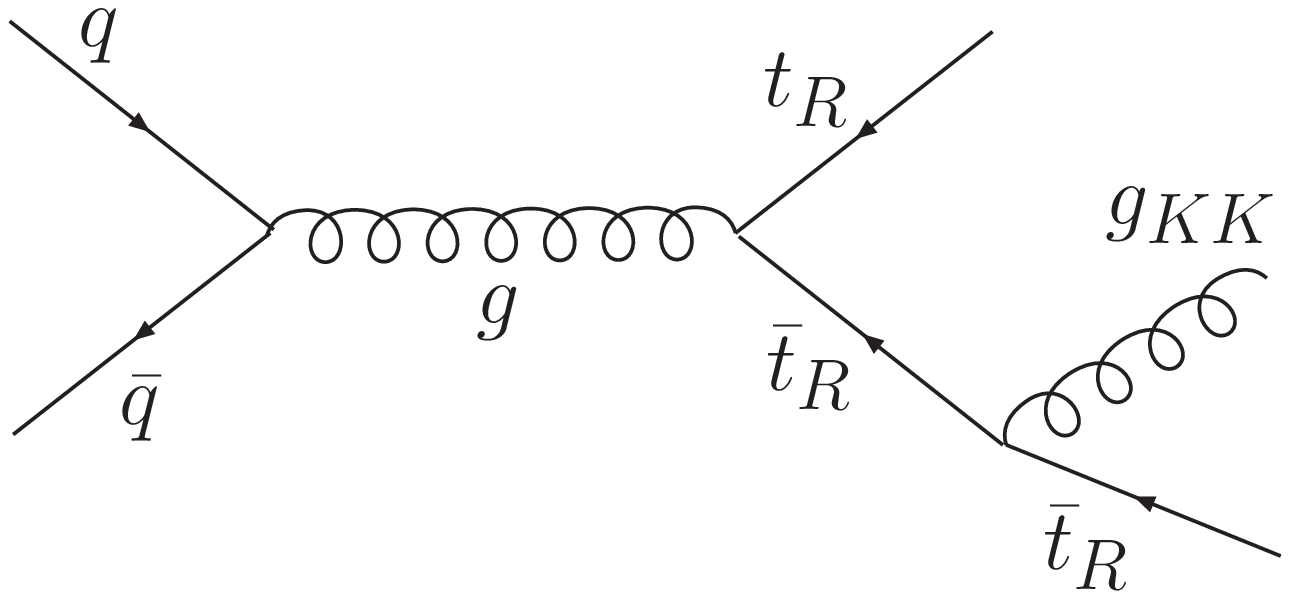}}

\hbox{(b)} \hbox{\includegraphics[scale=0.3,bb=101 470 495 723]{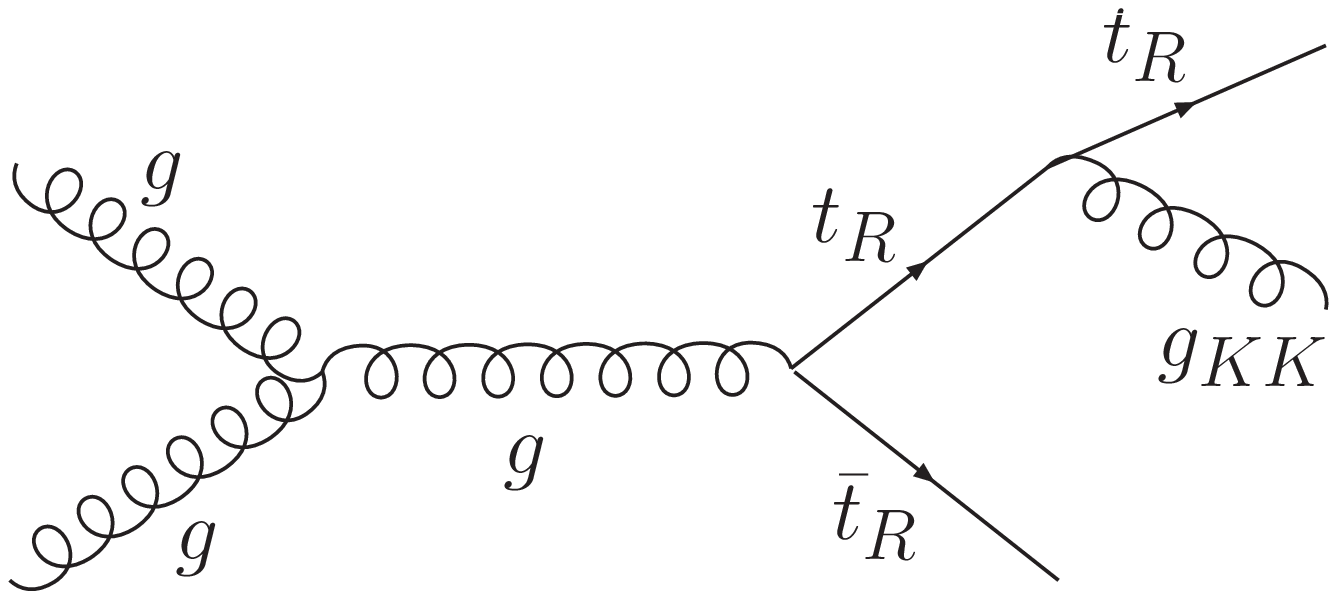}~\includegraphics[scale=0.3,bb=95 500 500 721]{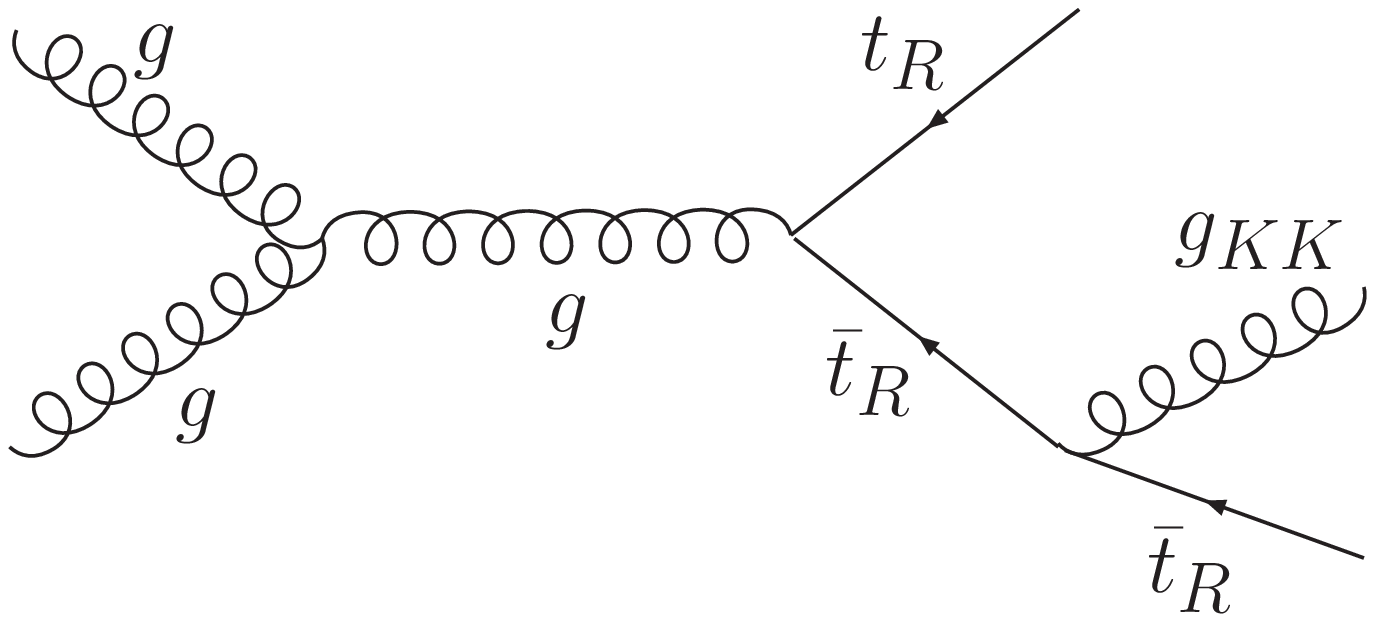}~\includegraphics[scale=0.3]{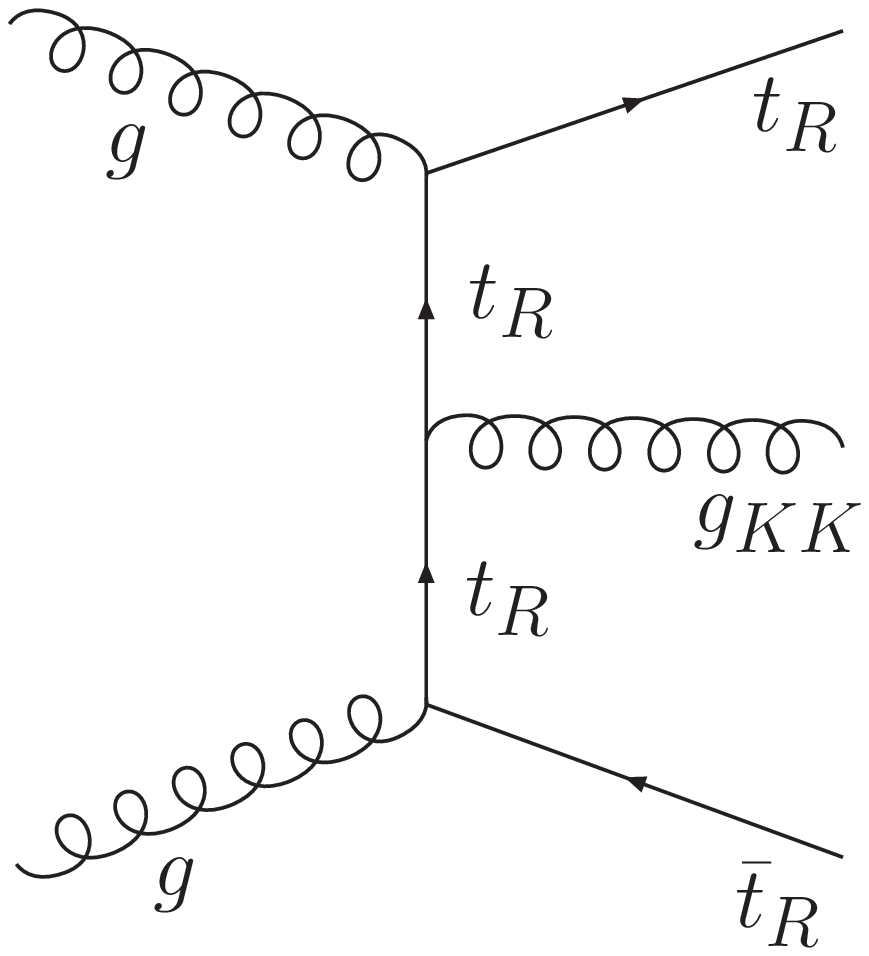}~\includegraphics[scale=0.3]{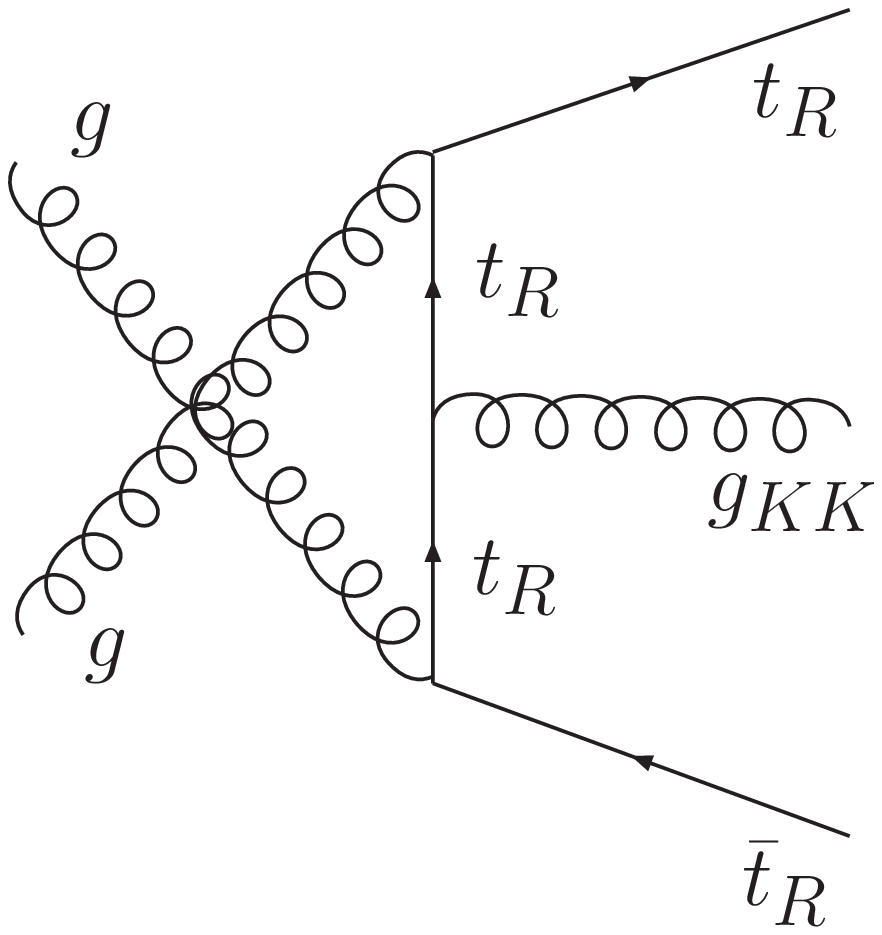}}

\hbox{\includegraphics[scale=0.3,bb=132 400 464 726]{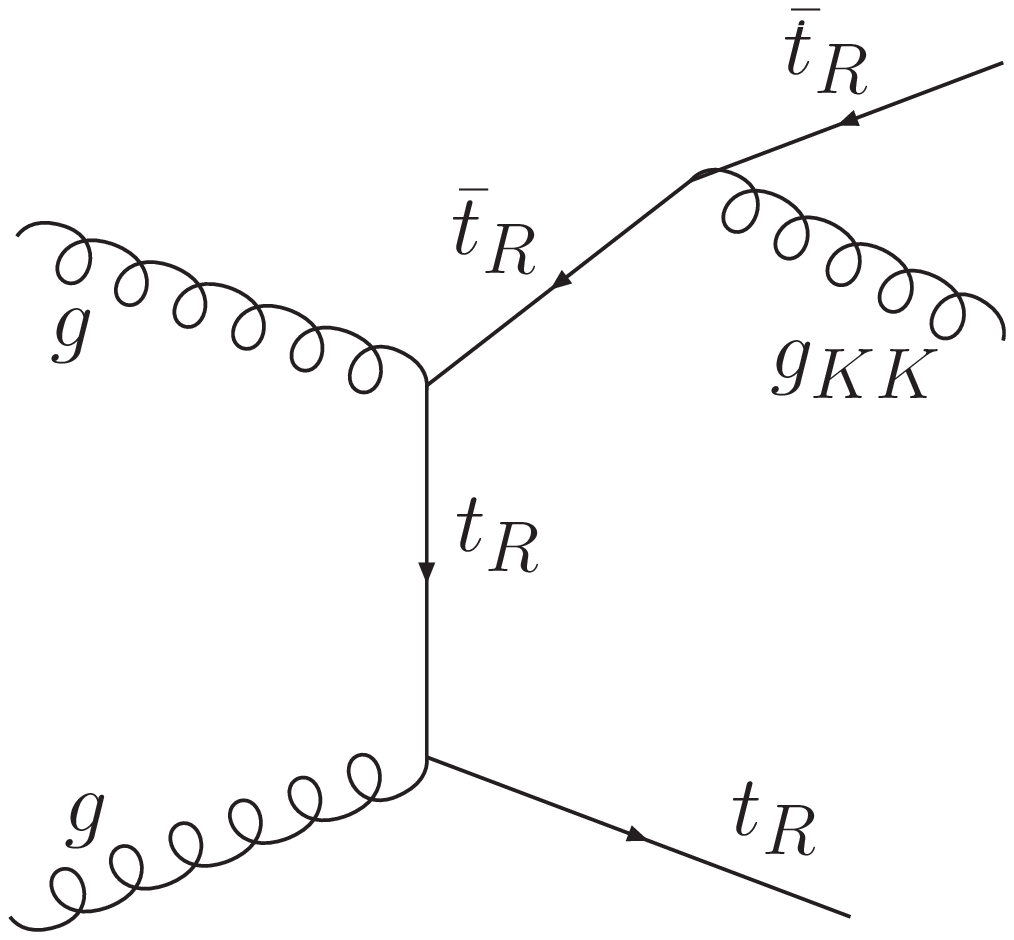}~\includegraphics[scale=0.3,bb=132 405 464 726]{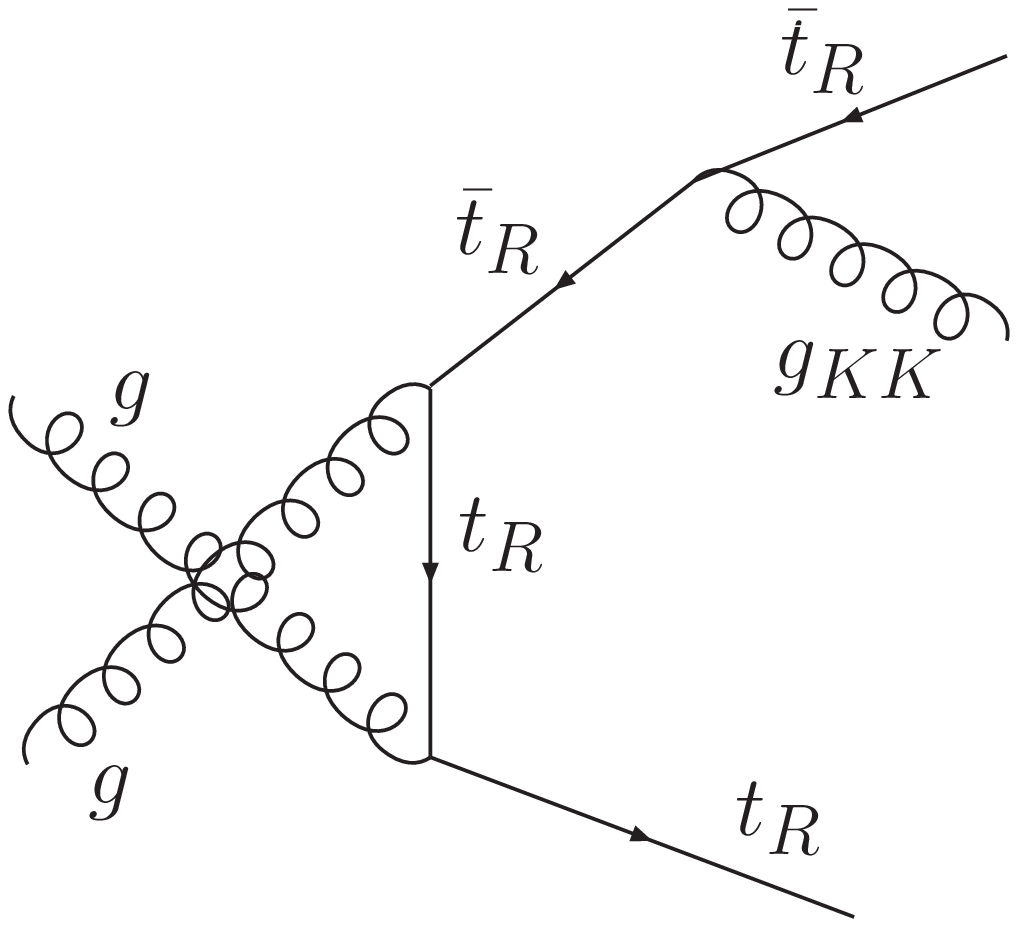}~\includegraphics[scale=0.3]{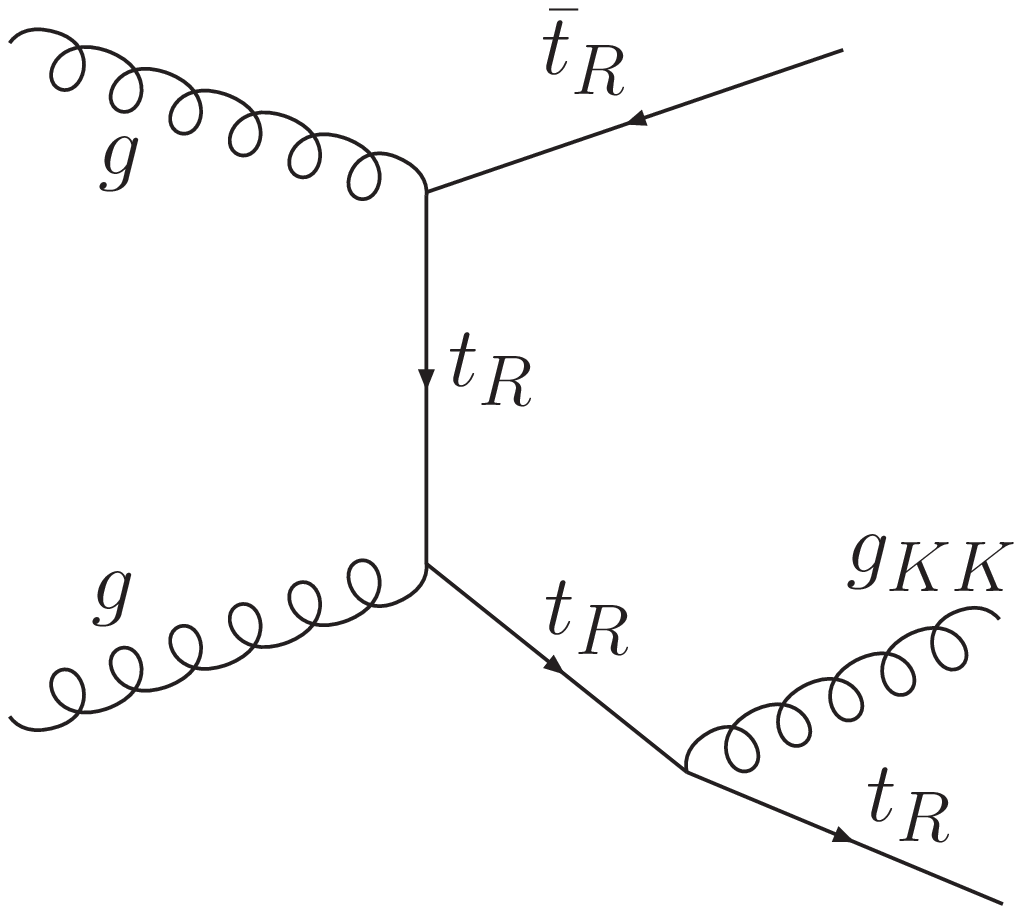}~~\includegraphics[scale=0.3]{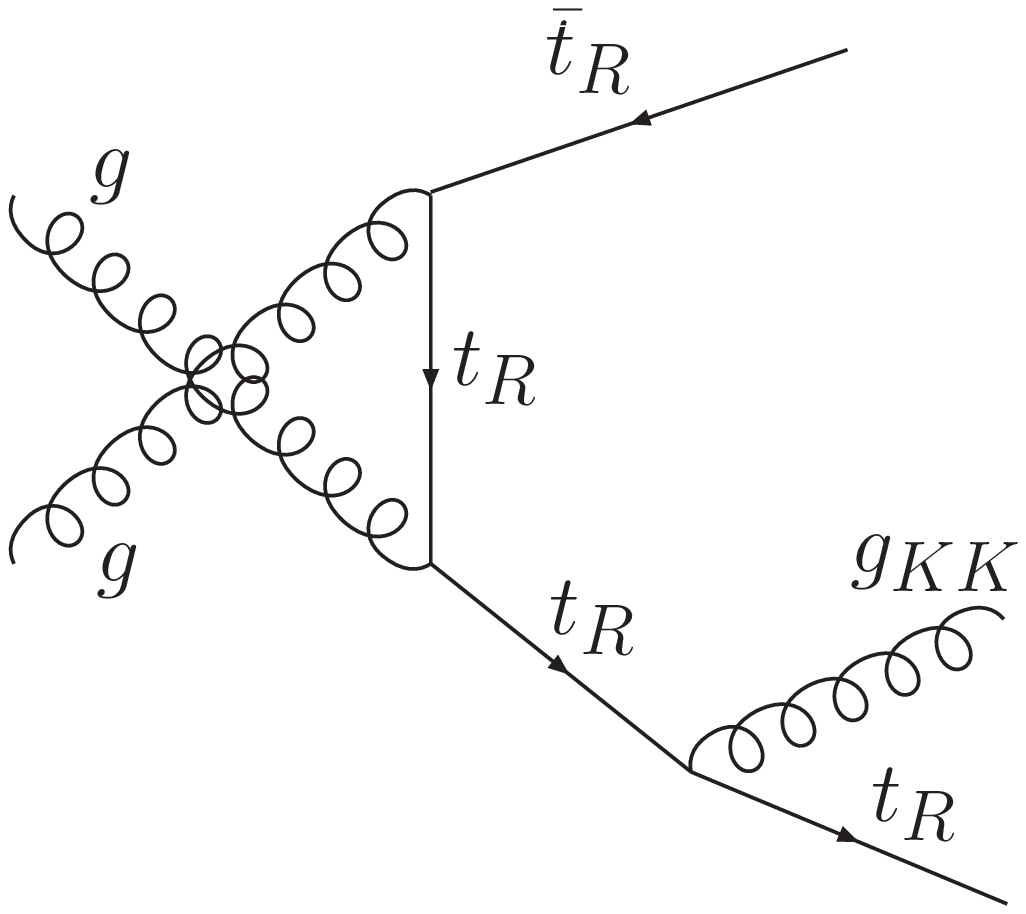}}

\caption{The Feynman diagrams for the processes: (a) $q \bar q \to g_{KK} t_R \bar t_R$ and, (b) $gg \to g_{KK} t_R \bar t_R$.}
\end{figure}%

For the signal kinematics, we have used that originally proposed by
Gottschalk and Sivers \cite{sivers} for three-jet production suitably
modified to take into account the fact that all three final-state particles
in our case are massive. In this description of the kinematics, the $z$-axis
of the co-ordinate system is chosen to be the direction of one of the
final-state particles rather than the initial beam axis. We choose
this particle (labelled $p_5$) as the $g_{KK}$. The momentum assignments
that we start with are:
\begin{eqnarray}
p_1 &:& {\sqrt{\hat s} \over 2}(1,\ \phantom{-}{\rm sin}\theta{\rm cos}\phi,\ 
\phantom{-}{\rm sin}\theta{\rm sin}\phi, {\rm cos}\theta) \nonumber \\
p_2 &:& {\sqrt{\hat s} \over 2}(1,\ - {\rm sin}\theta{\rm cos}\phi,\ 
- {\rm sin}\theta{\rm sin}\phi, {\rm cos}\theta) \nonumber \\
p_3 &:& {\sqrt{\hat s} \over 2} x_3 (1,\ \beta_3 {\rm cos}\theta_{35},\ 
\beta_3 {\rm sin}\theta_{35},\ 0) \nonumber \\
p_4 &:& {\sqrt{\hat s} \over 2} x_4 (1,\ \beta_4 {\rm cos}\theta_{45},\ 
\beta_4 {\rm sin}\theta_{45},\ 0) \nonumber \\
p_5 &:& {\sqrt{\hat s} \over 2} x_5 (1,\ \beta_5,\ 0,\ 0) .
\end{eqnarray}

In the above equation, $p_1$ and $p_2$ are the 4-momenta of the initial
partons, $p_3$ and $p_4$ are the 4-momenta of the $t$ and the $\bar t$
and $p_5$ is the 4-momentum of the KK gluon. The $\beta_i$'s are given
by:
\begin{equation}
\beta_i=\sqrt{1-{4m_i^2 \over x_i^2 \hat s}}
\end{equation}
Energy conservation implies
\begin{equation}
x_3+x_4+x_5=2,
\end{equation}
and, using 3-momentum conservation, one can get
\begin{eqnarray}
{\rm cos}\theta_{35} &=& {x_4^2\beta_4^2 -x_3^2\beta_3^2-x_5^2\beta_5^2
\over 2x_3x_5\beta_3\beta_5}
\nonumber \\
{\rm cos}\theta_{45} &=& {x_3^2\beta_3^2 -x_4^2\beta_4^2-x_5^2\beta_5^2
\over 2x_4x_5\beta_4\beta_5}
\end{eqnarray}

Using the above, all relevant momenta and angles may be constructed. For 
example, the transverse momentum of the $g_{KK}$ is given by
\begin{equation}
p_T (g_{KK})={\sqrt{\hat s} \over 2}x_5\beta_5\sqrt{{\rm cos}^2\theta+
{\rm sin}^2\theta {\rm sin}^2\phi}
\end{equation}
The kinematics for the decay of the $g_{KK}$ into a $t \bar t$ pair
is the standard two-particle decay kinematics. 

Defining the variables $\tau$ and $y_{\rm boost}$ through the equations:
\begin{eqnarray}
\tau &=& {\hat s \over s}=x_1x_2 \nonumber \\
x_1&=&\sqrt{\tau} e^{y_{\rm boost}} \nonumber \\
x_2&=&\sqrt{\tau} e^{-y_{\rm boost}} ,
\end{eqnarray}
the differential
cross-section for $g_{KK} t \bar t$ production assumes the form:
\begin{equation}
{d\sigma \over d\sqrt{\hat s} dy_{\rm boost} dx_3 dx_4 d\Omega} =
\int dx_1 dx_2
{\alpha_s^2 \Lambda_t^2 \tau \over 8 \pi \sqrt{\hat s}}\sum_{ij} 
{1 \over 1+\delta_{ij}} \biggl\lbrack f_i^{(a)}(x_1, Q^2) f_j^{(b)}(x_2, Q^2) 
\vert M_{ij}\vert^2 + i \leftrightarrow j \biggr\rbrack 
\end{equation}
where, $f^{(a)},\ f^{(b)}$ are the parton densities evaluated at the scale
$Q^2$, $\vert M_{ij} \vert^2$ is
the squared-matrix element and $\Lambda_t$ is the coupling of the KK gluon to
the $t_R$ and is given by $5 \sqrt{4 \pi \alpha_s}$. 

Since the $g_{KK}$ masses we are interested in are large, we expect 
the $t \bar t$ pair coming from the decay of the $g_{KK}$ to have large
momenta. The other $t \bar t$ pair is expected to have more moderate
values of momenta. This simple fact may allow one to enhance the quality
of the $g_{KK}$ signal over the QCD background. As a first guess, we choose
to put a lower cut of 300 GeV on the $p_T$ of the $t$ and the $\bar t$ 
coming from the decay of the $g_{KK}$ and a cut of 50 GeV on the each of
the other pair. We use these cuts to calculate the cross-section for
the associated production of the KK gluon with a $t \bar t$ pair.

\begin{figure}[htb]
\begin{picture}(4,6)
\put(0,0){\epsfig{file=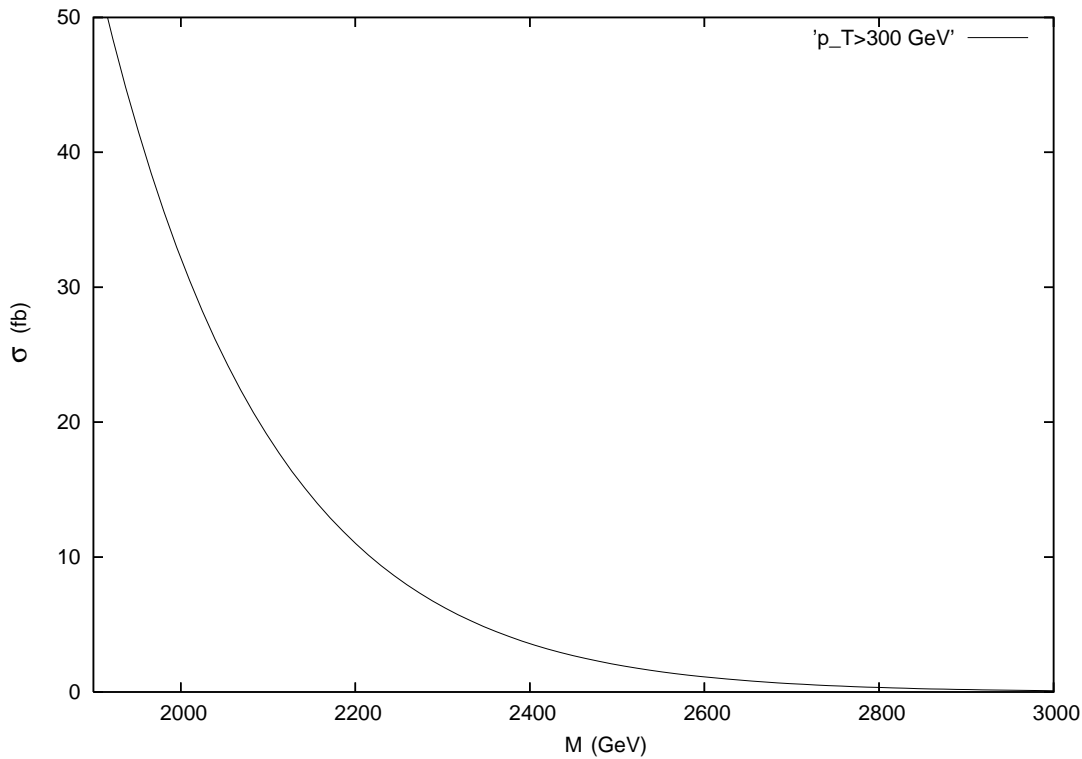, width=4in, angle=-90}}
\end{picture}
\vspace{10cm}
\caption{\it The cross-section for the production of a KK gluon in 
association with a $t \bar t$ pair at the LHC energy as a function
of the KK gluon mass and with $p_T$ cuts as described in the text. }  
\protect\label{fig2}
\end{figure}

In Fig.~2, we have plotted this cross-section as a function of the mass of
the KK gluon, $M$,
for $pp$ collisions at the LHC energy of $\sqrt{s}=14$~TeV. 
We have used the CTEQ4M densities \cite{cteq} and the parton distributions
are taken from PDFLIB \cite{pdflib}. For the QCD scale, we use $Q=\sqrt{\hat 
s}/2$. For this choice of parameters and cuts, we have used ALPGEN to 
compute the background and we find a background cross-section of  0.33 fb.
Assuming an integrated luminosity of 100 fb${}^{-1}$, we find from Fig. 2
that a significance ($ \equiv S/\sqrt{B}$) of 5 is obtained for $M=2790$ GeV.
The fact that the kinematic reach that this channel provides in searching
for the KK gluon at the LHC is of the same order of magnitude as allowed
by precision electroweak measurements is encouraging. Note that we have
made no attempt to optimise the significance of the signal and a more
judicious choice of cuts could conceivably help in increasing the reach.

\begin{figure}[htb]
\begin{picture}(4,6)
\put(0,0){\epsfig{file=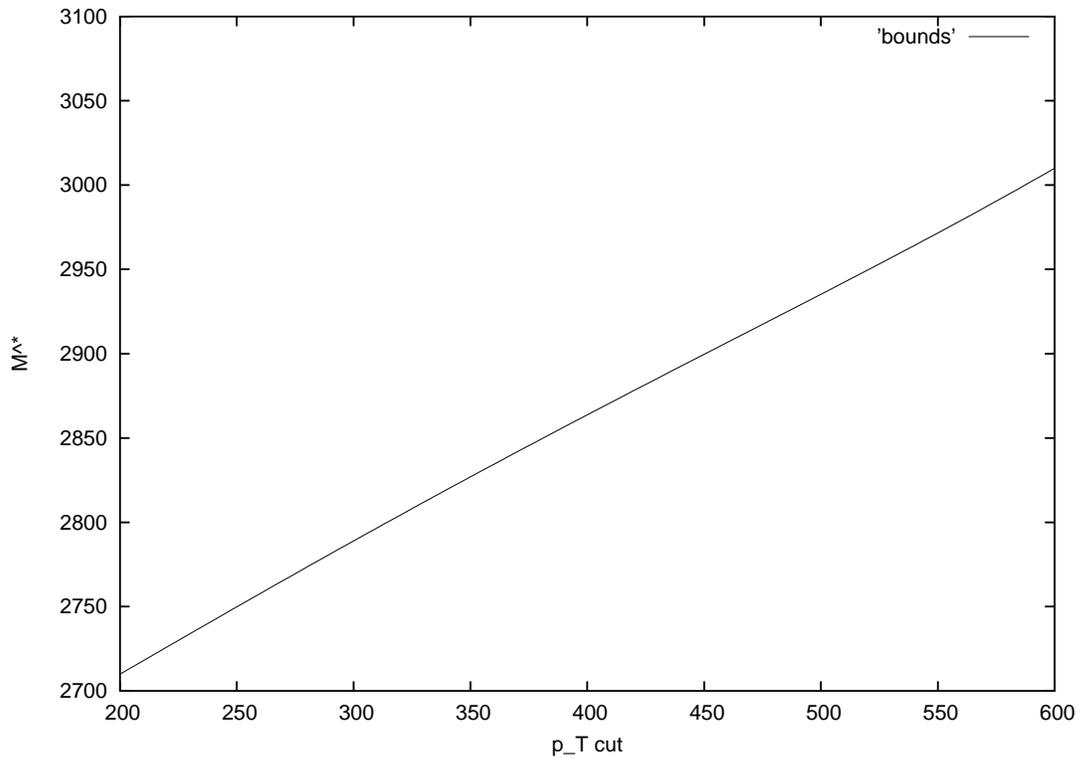, width=4in, angle=-90}}
\end{picture}
\vspace{10cm}
\caption{\it The reach, $M^*$, in KK gluon mass at the LHC
as a function of the $p_T$ cut. }  
\protect\label{fig3}
\end{figure}

Since the choice of 300 GeV for the value of the cut on the $p_T$ of the
top quarks coming from the decay of the KK gluon was only an educated guess,
we also studied the effect of changing this cut on the significance of
the signal. In Fig. 3, we have displayed the results of varying the cut on
$p_T$ assuming an integrated luminosity of 100 fb${}^{-1}$. 
For different values of the cut we have plotted the value, $M^*$, 
of the mass of the $g_{KK}$ for which a significance of 5 is obtained.
We find that changing the cut from 200 GeV to 600 GeV increases the reach
by about 300 GeV. But in choosing a larger $p_T$ cut one loses out on the 
number of events so that there are hardly a couple of background events 
at a $p_T$ cut of 600 GeV.
Therefore, one may have to optimise the cut by choosing it to be around
300 or 400 GeV which leaves us with a sizeable number of events.

The preferential coupling of $g_{KK}$ to $t_R$ as opposed to $t_L$ can
be exploited to increase the significance of this signal. The chiral coupling
of the $g_{KK}$ suggests that the polarization of the top quarks, studied
by looking at its decay products, can prove to be a very useful discriminator
between the signal and the background. However, in the present paper which
is based on a parton-level Monte Carlo study, we
have limited ourselves to studying the kinematic reach of the LHC in
the associated production process because the 4-top final state that we
have focussed on here is not going to be an easy final state to analyze
at the LHC experiments given the combinatorial backgrounds from this state
that would have to be dealt with to extract a realistic signal. We have
deferred a more detailed study of this signal after implementing it in
a hadron-level Monte Carlo. This will enable us to present various kinematic
distributions and top-polarization studies. Nevertheless, the results
presented in this paper are interesting enough to urge experimentalists
at the LHC to consider this process seriously.

\noindent{\sl Acknowledgements}:\\
One of us (K.S) would like to acknowledge fruitful discussions with 
Kaustubh Agashe, Abdelhak Djouadi, Riccardo Rattazzi and Bryan Webber.

 
\end{document}